\documentclass[10pt,conference]{IEEEtran}
\usepackage{multirow}
\usepackage{amsmath}
\usepackage[dvipsnames]{xcolor}
\usepackage{tikz}
\usepackage{listings}
\usepackage{fontawesome5}
\usepackage{tabularx}
\usepackage{booktabs}
\usepackage{amssymb} 
\usepackage{comment}
\usepackage{longfbox}
\usepackage{tikz}
\usepackage{pgfplots}
\usetikzlibrary{positioning,shapes,shadows,arrows}
\usepackage[style=base]{subcaption}
\usepackage{graphicx}
\usepackage{url}
\usepackage{hyperref}
\usepackage{cleveref}
\usepackage{tcolorbox}

\title{Extracting Formal Specifications from Documents Using LLMs for Automated Testing}

\author{
\IEEEauthorblockN{
Hui Li\textsuperscript{1},
Zhen Dong\textsuperscript{1},
Siao Wang\textsuperscript{1},
Hui Zhang\textsuperscript{1},
Liwei Shen\textsuperscript{1},
Xin Peng\textsuperscript{1},
Dongdong She\textsuperscript{2}
}
\IEEEauthorblockA{\textsuperscript{1}Fudan University, Shanghai, China \\
Email: \{22210240023, 22110240039, 24210240401\}@m.fudan.edu.cn \{zhendong, shenliwei, pengxin\}@fudan.edu.cn}
\IEEEauthorblockA{\textsuperscript{2}The Hong Kong University of Science and Technology, Hong Kong, China \\
Email: dongdong@cse.ust.hk}
}

\author{
\IEEEauthorblockN{
Hui Li\textsuperscript{1},
Zhen Dong\textsuperscript{1,\dag},
Siao Wang\textsuperscript{1},
Hui Zhang\textsuperscript{1},
Liwei Shen\textsuperscript{1},
Xin Peng\textsuperscript{1},
Dongdong She\textsuperscript{2}
}
\IEEEauthorblockA{\textsuperscript{1}Fudan University, Shanghai, China \
Email: {22210240023, 22110240039, 24210240401}@m.fudan.edu.cn \\ {zhendong, shenliwei, pengxin}@fudan.edu.cn}
\IEEEauthorblockA{\textsuperscript{2}The Hong Kong University of Science and Technology, Hong Kong, China \
Email: dongdong@cse.ust.hk}
\IEEEauthorblockA{\textsuperscript{\dag}Corresponding author}
}

\definecolor{lightgreen}{RGB}{0.9, 1, 0.9}
\definecolor{myolive}{RGB}{128,128,0}
\definecolor{myhulk}{RGB}{0,128,0}
\definecolor{myteal}{RGB}{0,128,128}

\definecolor{specbg}{RGB}{255,250,205}
\definecolor{instructionbg}{RGB}{255,240,245}
\definecolor{codebg}{RGB}{255,248,220}
\definecolor{cotbg}{RGB}{255,248,215}
\definecolor{reqbg}{RGB}{255,243,210}

\definecolor{iobg}{RGB}{255,238,205}
\definecolor{unknownbg}{RGB}{255,199,206}

\pgfplotsset{compat=1.18}
\begin{document}

\maketitle

\begin{abstract}
Automated testing plays a crucial role in ensuring software security. It heavily relies on formal specifications to validate the correctness of the system behavior. 
However, the main approach to defining these formal specifications is through manual analysis of software documents, which requires a significant amount of engineering effort from experienced researchers and engineers. Meanwhile, system update further increases the human labor cost to maintain a corresponding formal specification, making the manual analysis approach a time-consuming and error-prone task. 

Recent advances in Large Language Models (LLMs) have demonstrated promising capabilities in natural language understanding.
Yet, the feasibility of using LLMs to automate the extraction of formal specifications from software documents remains unexplored. 
We conduct an empirical study by constructing a comprehensive dataset comprising 603 specifications from 37 documents across three representative open-source software. We then evaluate the most recent LLMs' capabilities in extracting formal specifications from documents in an end-to-end fashion, including GPT-4o, Claude, and Llama. 

Our study demonstrates the application of LLMs in formal specification extraction tasks while identifying two major limitations: specification oversimplification and specification fabrication. We attribute these deficiencies to the LLMs' inherent limitations in processing and expressive capabilities, as well as their tendency to fabricate fictional information. 
Inspired by human cognitive processes, we propose a two-stage method, annotation-then-conversion, to address these challenges.  
Our method demonstrates significant improvements over the end-to-end method, with a 29.2\% increase in the number of correctly extracted specifications and a 14.0\% improvement in average accuracy. In particular, our best-performing LLM achieves an accuracy of 71.6\%. 
\end{abstract}

\section{Introduction}

Automated testing is an important software testing technique to discover vulnerabilities \cite{rafi2012benefits, kasurinen2010software}. 
A critical component of automated testing is the test oracle that can tell whether the output of a system under test is correct \cite{barr2014oracle}. 
In practice, the accuracy of a test oracle is often determined by the formal specifications of the system under test \cite{ibrahimzada2022perfect, staats2011programs}. 
It is quite challenging to automatically derive formal specifications from software documents\cite{le2018deep}. 

Currently, the main method for defining these formal specifications is through manual analysis of software documents\cite{zhai2020c2s, al2024hermes}. 
However, a significant amount of engineering effort is required from experienced researchers or engineers who have sufficient domain knowledge and experience in the expected system behaviors. This time-consuming manual analysis further drastically increases the human labor cost in automated testing. 
Moreover, as modern software systems are constantly evolving, the corresponding formal specifications need frequent updates, which require a great deal of manual analysis effort.  
To address this issue, an automated method is urgently needed to generate formal specifications. Because it can minimize manual intervention, improve efficiency, and reduce human labor costs in automated testing.

Recent advances in large language models (LLMs) have demonstrated their impressive ability to comprehend and generate text across various domains \cite{zhao2023survey, chang2024survey}. The latest breakthrough in LLMs indicates great potential for automating the extraction of formal specifications from software documentation\cite{xu2023large}. Therefore, we investigate the feasibility of using LLMs to extract specifications from documents in an end-to-end fashion, with a primary focus on addressing the following research question:

\noindent \textbf{RQ1}: To what extent can the LLM extract formal specifications from the software document in an end-to-end fashion?

This study comprises two primary components: the development of the dataset and the evaluation of our LLM-based formal specification extraction technique. 
To construct the dataset, we use three representative open-source software with elaborate behavior requirements and constraints in the documents, including ArduPilot \cite{ardupilot-website} and PX4 \cite{px4-website} for unmanned aerial vehicle (UAV) flight control and Autoware \cite{autoware-website} for autonomous driving.  We meticulously collected 37 specification-related documents with an average length of 2,966 words across three projects.
To obtain the ground truth of the formal specifications, we ask two domain experts in formal methods and software testing to independently extract and cross-validate temporal logic \cite{pnueli1977temporal} specifications from these collected documents, producing a total of 603 validated specifications.
We evaluate the effectiveness of the end-to-end method using three state-of-the-art language models: GPT-4o \cite{openai_gpt4o}, Claude-3.5-Sonnet \cite{Anthropic_Claude35Sonnet}, and Llama-3.1-405B \cite{Meta_Llama31}. We craft a straightforward prompt using the roleplaying prompt engineering technique to unlock LLM's potential in specification extraction.

Our study reveals not only the promising results but also the limitations of using LLMs to extract the formal specifications in an end-to-end fashion. 
The study showed exciting results that Claude-3.5-Sonnet achieved the highest accuracy at 51.7\%, followed by GPT-4o at 47.1\%, and LLama-3.1-405B at 45.4\%. 
However, these LLMs struggled with documents that contained a large number of formal specifications. For example, the largest document can contain up to 44 formal specifications. Even the best-performing LLM only correctly identified 15 specifications, achieving a low accuracy of 34.1\%.

Furthermore, a thorough analysis of extracted formal specifications revealed two major limitations of the end-to-end method. First, the LLMs tend to produce overly simplistic boundary conditions and reduce complex specifications to redundant and simplistic temporal logic formulas. Second, the LLMs will make up fake formal specifications by introducing details not present in the software documentation, resulting in plausible but factually wrong specifications.

We attribute these deficiencies to the inherent limitations of LLM, including their limited processing and expressive capabilities \cite{liu2024lost, feng2024towards}, as well as their propensity to hallucination \cite{ji2023survey}.
The task of extracting specifications from documents places a significant burden on LLMs, requiring them to process vast amounts of text and articulate complex specifications simultaneously, which can exceed their capability limit. Moreover, the hallucination of LLMs also means that they cannot guarantee reliable formal specifications with respect to the software document when generating specifications.

To overcome these challenges, we propose an annotation-then-conversion method, which breaks down the specification extraction task into two manageable subtasks: sentence annotation and temporal logic conversion. By mimicking human cognitive processes and decomposing the task into smaller subtasks, this method reduces the demands on LLMs' processing and expressive capabilities. Additionally, our method enables effective fact-checking by generating verifiable pairs of sentences and specifications, thereby minimizing the impact of hallucination.

We evaluate annotation-then-conversion method on the constructed dataset through two research questions:

\noindent \textbf{RQ2}: Can the annotation-then-conversion technique improve the LLM's ability of formal specification extraction?

\noindent \textbf{RQ3}: How do LLMs perform on the task of temporal logic conversion compared to the state-of-the-art method in the pre-LLM era?

Our experimental results demonstrate the effectiveness of annotation-then-conversion method. Compared to the end-to-end method, the method achieved a notable 14.0\% increase in average accuracy and a 29.2\% increase in the number of correct extracted specifications. 
Our results confirm the effectiveness of the proposed method. 
We open-source the experiment data and implementations in an Anonymous repository at \href{https://github.com/lhorse010/llm_specificaiton_extraction.git}{https://github.com/lhorse010/llm\_specificaiton\_extraction}.

In summary, this paper makes the following key contributions:
\begin{itemize}
    \item Our empirical study revealed the limitations of end-to-end method in specification extraction, highlighting the need for a more effective method.

    \item We proposed a novel two-stage specification extraction method, annotation-then-conversion, which achieved a significant improvement in accuracy.

    \item We constructed a dataset comprising 37 documents and 603 verified specifications, which will facilitate further research.
\end{itemize}

\section{Empirical Study}
\label{sec:empirical_study}

\subsection{Resarch Questions}
\noindent \textbf{RQ1}: To what extent can the LLM extract formal specifications from the software document in an end-to-end fashion?

RQ1 aims to investigate how well LLMs can extract specifications in temporal logic formulas from documents. It is a straightforward method to feed documents into LLMs and provide suitable prompts to extract specifications in temporal logic.  
We apply this method and evaluate LLM's ability to extract specifications in temporal logic from the raw documents.

\subsection{Subjects and Dataset}
\subsubsection{LLM Selection}
We select LLMs for evaluation based on three key criteria: popularity, diversity, and capability. Specifically, we consider LLMs that are widely used, developed by different organizations, and include a mix of open-source and close-source options. Additionally, we prioritize LLMs with advanced capability.

Our selection includes two state-of-the-art close-source LLMs: GPT-4o by OpenAI and Claude-3.5-Sonnect by Anthropic. Notably, we excluded the o1-preview version of GPT-4 from consideration due to its high computational latency and high API costs. To assess the capabilities of open-source LLMs, we also include a leading open-source model, Llama-3.1-405B by Meta.

\subsubsection{Software Selection}
We selected open-source software for our study based on three primary criteria: availability of behavioral requirements, popularity, and diversity. Specifically, we focused on popular GitHub projects with extensive behavioral requirements and constraints and chose software from different communities to increase the generalization of our findings. Our selected software includes ArduPilot and PX4, two widely-used UAV flight control software with comprehensive flight behavior requirements and constraints, as well as Autoware, a leading autonomous driving framework with detailed behavioral requirements for driving scenarios.

\subsubsection{Document Selection}
We prioritize documents relevant to control modules for each software project, as they play a crucial role in determining the behavior of the software and provide valuable insights into the system's specifications and behavior. This initial selection yields 25 documents for Ardupilot, 18 for PX4, and 28 for Autoware.

To further refine our dataset, we favor documents with minimal multi-modal content, such as figures and videos, to ensure that we can focus on the textual specifications and avoid potential ambiguities. By excluding documents with extensive multi-modal information, we obtain a final dataset of 21 documents for Ardupilot, 11 for PX4, and 5 for Autoware.

\subsubsection{Document Preprocessing}
\begin{figure}[!htbp]
    \centering
    \includegraphics[width=\linewidth]{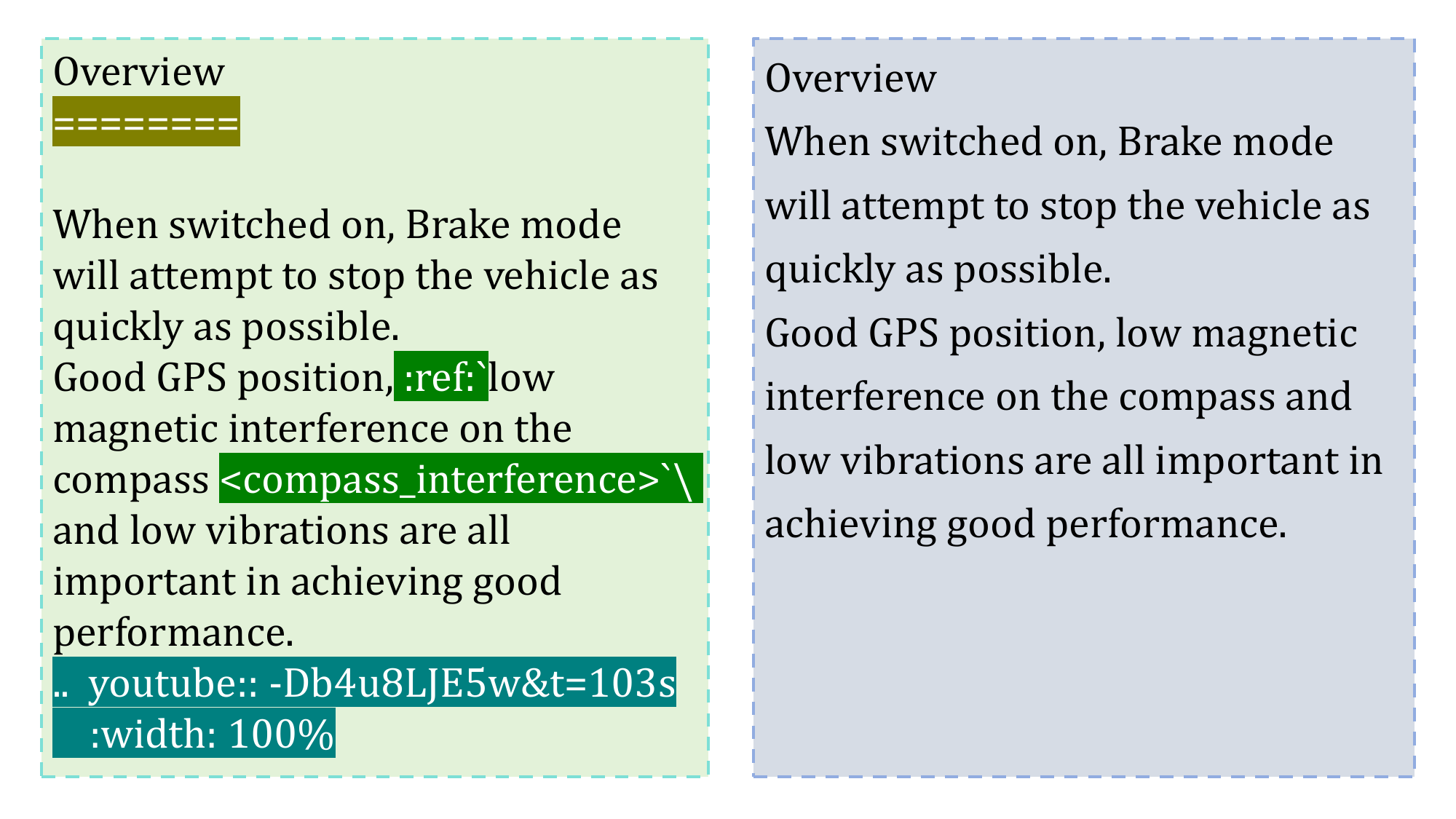}
    \caption{Document Preprocessing Illustration: The original document file (left) and the preprocessed document (right) are shown. We remove syntax markup as in \colorbox{myolive}{\textcolor{white}{olive}}, references as in \colorbox{myhulk}{\textcolor{white}{green}}, and multi-modal elements as in \colorbox{myteal}{\textcolor{white}{teal}}.}
    \label{fig:doc_preprocess}
\end{figure}

As \Cref{fig:doc_preprocess} shows, during the preprocessing phase, we initially eliminated video and image content from the multi-modal documentation. 
Subsequently, we removed formatting elements, including Markdown and reStructuredText (RST) syntax, to convert the documentation into plain text format. 
Finally, we segmented the documents into individual sentences and restructured them into a standardized format that includes paragraph titles followed by their corresponding textual content.

On average, the documents in our dataset contain 2966 words, with the longest document comprising 9938 words.

\subsubsection{Ground Truth Obtain}
To establish a reliable ground truth dataset, we enlisted the expertise of two researchers in formal methods and software testing. We asked them to independently examine the documents and extract specifications in the form of temporal logic formulas. Once they had completed their extractions, we had them exchange their results and perform a cross-check to ensure accuracy. Only formulas that were unanimously approved by both experts were added to the ground truth dataset.

Through this rigorous process, we obtained a total of 603 specifications from 37 documents.

\subsection{Study Method for RQ1}

\begin{figure}
\centering
\begin{tikzpicture}[node distance=0.2cm]

\tikzstyle{instruction} = [rectangle, rounded corners, fill=instructionbg,  text width=8cm, align=left, font=\small, inner sep=5pt]
\tikzstyle{objective} = [rectangle, rounded corners, fill=codebg,  text width=8cm, align=left, font=\small, inner sep=5pt]
\tikzstyle{io} = [rectangle, rounded corners, fill=iobg,  text width=8cm, align=left, font=\small, inner sep=5pt]

\node[instruction] (roleplay) {
   \textbf{Roleplaying}\\
    You are an expert in Temporal Logic (TL) with years of experience in formal verification and testing. 
};

\node[objective, below=of roleplay] (obj) {
   \textbf{Objective}\\
   Extract specifications that the vehicle needs to meet from the document. Then express them using Temporal Logic (TL) formulas with the following symbols:
    
    \textbf{Logical operators}:
    \begin{itemize}
    \item $\neg$ (negation)
    \item $\lor$ (or)
    \item $\land$ (and)
    \item $\rightarrow$ (implies)
    \end{itemize}
    
    \textbf{Temporal modal operators}:
    \begin{itemize}
    \item X (next)
    \item U (until)
     \item G (globally)
     \item  F (finally)
    \end{itemize}
};

\node[io, below=of obj](io) {
\textbf{Output Format}
\begin{lstlisting}[basicstyle=\ttfamily\scriptsize, breaklines=true]
 {
    "specifications":[
        {
            "formula": ...,
            "explanation": ...
        },...
    ] 
}   
\end{lstlisting}
};

\end{tikzpicture}
\caption{End-to-End Extraction Prompt Template. This template consists of three key components: (1) Roleplaying technique, which guides the LLM to emulate an expert and tailor its response to the extraction task; (2) Objective statement, which clearly defines the goal of the extraction task; and (3) Output format specification, which ensures that the result conforms to the desired format.}
\label{fig:prompt-end-to-end}
\end{figure}

To investigate RQ1, we extracted specifications from documents using a roleplaying prompt template (shown in \Cref{fig:prompt-end-to-end}), guiding the LLM to act as a formal verification expert during the task execution. A comprehensive analysis was then conducted to evaluate the LLMs' performance, capabilities, and inherent limitations in handling specification extraction tasks.

The experimental framework was implemented on the Poe Platform using three distinct LLMs:
\begin{itemize}
    \item Claude-3.5-Sonnet (closed-source, accessed via Poe-official robot)
    \item GPT-4o (closed-source, accessed via Poe-official robot)
    \item Llama-3.1-405B-T (open-source, operated by together AI through the Poe Platform)
\end{itemize}

To mitigate the impact of stochastic variations in LLM outputs, we executed each LLM 3 times per input document, and all unique results were aggregated for subsequent evaluation. The validation process involved two domain experts who assessed each extracted specification. A specification was considered valid only if it satisfied two criteria: syntactic correctness and semantic alignment with the ground truth dataset.

Our performance evaluation framework employed two fundamental metrics:
\subsubsection{Accuracy} The Accuracy (ACC) measures how many right specifications(R) in the result match with the ground truth. It is defined as:

\[
ACC = \frac{|R|}{|GT|}
\]
 
A higher Accuracy indicates the LLM is more capable of extracting specifications from documents.

\subsubsection{False Positive}
The False Positive (FP) measures how many specifications in the result(RS) are fabricated or contain mistakes (W). It is defined as:
\[
FP = \frac{|W|}{|RS|}
\]
 
A lower FP indicates a higher reliability of the LLM in the specification extraction task,
a higher FP means the LLM is more likely to be hallucinated or incapable of tackling temporal logic.

\section{Study Results And Analysis}
\label{sec:study_analysis}
\subsection{Study Result}

\begin{table}[!ht]
    \centering
    \caption{Performance evaluation of specification extraction using end-to-end method across three LLMs: Claude, GPT-4o, and Llama. Results show extracted specifications from 37 documents(21 ArduPilot, 11 PX4, and 5 Autoware), where \textbf{r} represents correct extractions and \textbf{w} indicates wrong ones. The evaluation metrics, shown in the bottom two rows, include accuracy and false positive rate.}
    \label{tab:end_to_end_result}
    \scriptsize
    \begin{tabular}{l|cc|cc|cc|c}
    \toprule
        \multirow{2}{*}{\textbf{Document}} & \multicolumn{2}{c|}{\textbf{Claude}} & \multicolumn{2}{c|}{\textbf{GPT-4o}} & \multicolumn{2}{c|}{\textbf{Llama}}& \textbf{Ground} \\
        \cline{2-7}
        & \textbf{r} & \textbf{w} & \textbf{r} & \textbf{w} & \textbf{r} & \textbf{w} & \textbf{Truth} \\ 
        \midrule
        AP:Airmode & 6 & 0 & 7 & 0 & 6 & 0 & 8 \\ 
        AP:Auto & 10 & 2 & 14 & 6 & 13 & 8 & 29 \\ 
        AP:Brake & 7 & 1 & 3 & 1 & 6 & 1 & 8 \\ 
        AP:Circle & 15 & 0 & 16 & 0 & 16 & 0 & 25 \\ 
        AP:Drift & 10 & 2 & 7 & 2 & 9 & 1 & 14 \\ 
        AP:Flip & 7 & 1 & 7 & 7 & 6 & 4 & 9 \\ 
        AP:FlowHold & 7 & 1 & 4 & 5 & 4 & 3 & 8 \\ 
        AP:Follow & 5 & 8 & 8 & 4 & 10 & 7 & 12 \\ 
        AP:Guided & 8 & 4 & 8 & 6 & 9 & 3 & 27 \\ 
        AP:Heli\_Autorotate & 7 & 4 & 14 & 2 & 2 & 7 & 31 \\ 
        AP:Land & 9 & 1 & 6 & 2 & 8 & 1 & 11 \\ 
        AP:Loiter & 8 & 2 & 7 & 2 & 9 & 4 & 15 \\ 
        AP:PosHold & 8 & 1 & 5 & 1 & 7 & 1 & 11 \\ 
        AP:RTL & 15 & 1 & 13 & 4 & 15 & 4 & 44 \\ 
        AP:Simple & 6 & 4 & 7 & 1 & 8 & 1 & 19 \\ 
        AP:SmartRTL & 12 & 1 & 11 & 3 & 12 & 1 & 20 \\ 
        AP:Sport & 4 & 0 & 4 & 1 & 5 & 0 & 7 \\ 
        AP:Stabilize & 11 & 1 & 10 & 4 & 9 & 6 & 14 \\ 
        AP:SysID & 3 & 0 & 2 & 0 & 3 & 1 & 3 \\ 
        AP:Throw & 9 & 0 & 9 & 1 & 6 & 3 & 19 \\ 
        AP:Turtle & 7 & 1 & 3 & 2 & 5 & 1 & 12 \\ 
        PX4:Position & 9 & 0 & 5 & 0 & 5 & 0 & 20 \\ 
        PX4:Position Slow & 10 & 3 & 9 & 0 & 5 & 8 & 23 \\ 
        PX4:Altitude & 11 & 1 & 7 & 2 & 9 & 2 & 17 \\ 
        PX4:Stabilized & 10 & 4 & 5 & 0 & 6 & 2 & 16 \\ 
        PX4:Acro & 4 & 3 & 2 & 0 & 3 & 0 & 5 \\ 
        PX4:Hold & 9 & 2 & 8 & 4 & 5 & 2 & 13 \\ 
        PX4:Return & 10 & 3 & 10 & 6 & 11 & 4 & 22 \\ 
        PX4:Mission & 13 & 2 & 13 & 2 & 10 & 4 & 39 \\ 
        PX4:Takeoff & 8 & 2 & 7 & 2 & 7 & 2 & 11 \\ 
        PX4:Land & 11 & 0 & 7 & 2 & 5 & 2 & 13 \\ 
        PX4:Orbit & 10 & 0 & 12 & 3 & 14 & 3 & 27 \\ 
        AW:Blind Spot & 4 & 1 & 4 & 0 & 4 & 1 & 5 \\ 
        AW:Traffic Light & 8 & 2 & 9 & 1 & 6 & 0 & 9 \\ 
        AW:Detection Area & 7 & 0 & 5 & 2 & 6 & 0 & 8 \\ 
        AW:No Drivable Lane & 3 & 0 & 4 & 0 & 5 & 0 & 8 \\ 
        AW:Out of Lane & 11 & 4 & 12 & 5 & 5 & 1 & 21 \\ 
        \midrule
        \textbf{Sum} & 312 & 62 & 284 & 83 & 274 & 88 & 603 \\ 
        \midrule
        \textbf{Accuracy} & \multicolumn{2}{c|}{51.7\%} & \multicolumn{2}{c|}{47.1\%} & \multicolumn{2}{c|}{45.4\%} & - \\ 
        \textbf{False Positive} & \multicolumn{2}{c|}{16.6\%} & \multicolumn{2}{c|}{22.6\%} & \multicolumn{2}{c|}{24.3\%} & - \\
        \bottomrule
    \end{tabular}
\end{table}

\Cref{tab:end_to_end_result} compares how well different LLMs (Claude, GPT-4o, and Llama) perform at extracting specifications using the end-to-end method. The data covers multiple software systems: ArduPilot (21 documents), PX4 (11 documents), and Autoware (5 documents). For each document, the table shows the number of correctly extracted specifications (\textbf{r}) and incorrectly extracted specifications (\textbf{w}), compared against a ground truth. The bottom rows summarize the overall performance with accuracy and false positive rates.

The overall performance metrics demonstrate moderate capability in specification extraction across all three LLMs. Specifically, the LLMs extracted 312, 284, and 274 specifications for Claude, GPT-4, and Llama, respectively. Beyond the challenge of high false positive rates stemming from inherent limitations (such as hallucinations) of LLM, the accuracy rates reached above 45\% for all tested LLMs. Claude achieves relatively better performance with an accuracy of 51.7\%, followed by GPT-4 (47.1\%) and Llama (45.4\%), indicating that these LLMs can successfully extract about half of the specifications from the documentation. This performance level suggests that LLMs have potential in automated specification extraction tasks, though there is still considerable room for improvement.

However, significant challenges remain, particularly when handling modules with complex behavior, such as AP:Auto, AP:RTL, and PX4:Mission, resulting in considerable gaps between their extracted specifications and the ground truth. For instance, in the AP:RTL document, while the ground truth contains 44 specifications, even the best-performing LLM identified at most 15 correct specifications, achieving an accuracy of merely 34.1\%. These results reveal three key limitations in end-to-end specification extraction: insufficient capability in processing long documents, tendency to hallucinate non-existent details, and oversimplification of complex requirements. The current performance suggests the need to address these specific challenges to improve accuracy.

\subsection{Worse Case Analysis}
\subsubsection{Specification Oversimplification}
Our evaluation of the end-to-end specification extraction method revealed a significant shortcoming: it tends to generate overly simplistic specifications that do not adequately capture the complexity of system requirements. This limitation is evident in two key areas: an overemphasis on basic boundary conditions and the breakdown of requirements within sentences into excessively simplistic components.

The method exhibits a notable bias towards extracting simple boundary limits, typically in the form of threshold checks, ``$\text{G}(\text{value} \leq \text{threshold})$", such as ``$\text{G}(\text{moving\_distance} \leq \text{MAX\_DIST})$". While these constraints are essential, the method's focus on them is disproportionate, leading to inadequate capture of more complex system behaviors. This imbalance shows that the method has a significant limitation in capturing all system requirements, and more comprehensive extraction capabilities are needed.

\begin{figure}[!htbp]
\begin{tcolorbox}[
    title=\textbf{Example of Specification Oversimplification},
    colback=pink!10,
    colframe=orange!60!brown!60,
    boxrule=2pt,
    width=\columnwidth]

\textbf{Text in the Document:}

``It will climb or descend at up to 2.5m/s"

\textbf{Generated Oversimplification:}

$\mathit{G}(\mathit{climbing} \rightarrow \mathit{G}(\mathit{vertical\_velocity} \leq \mathit{2.5m/s})$

$\mathit{G}(\mathit{descending} \rightarrow \mathit{G}(\mathit{vertical\_velocity} \geq \mathit{-2.5m/s})$
 
\end{tcolorbox}

\caption{Example of specification oversimplification: LLMs may break down a single requirement within a sentence into multiple naive formulas.}
\label{fig:example_of_specification_oversimplification}
\end{figure}

\Cref{fig:example_of_specification_oversimplification} shows that LLM will break down behavioral requirements within sentences into smaller components, leading to unnecessary duplication. 
Consider the straightforward requirement \textit{``It will climb or descend at up to 2.5m/s".} 
The LLM separates this into two distinct temporal logic constraints:
\begin{itemize}
    \item $\mathit{G}(\mathit{climbing} \rightarrow \mathit{G}(\mathit{vertical\_velocity} \leq \mathit{2.5m/s})$
    \item $\mathit{G}(\mathit{descending} \rightarrow \mathit{G}(\mathit{vertical\_velocity} \geq \mathit{-2.5m/s})$
\end{itemize}

While technically correct, this separation is unnecessary and only serves to increase complexity and reduce the system's ability to process additional requirements.

The output capacity constraints inherent to LLM end-to-end methods result in these simplified specifications dominating the output space, potentially excluding more sophisticated requirements.

\subsubsection{Specification Fabrication}

Our analysis of LLM-generated specifications exposed a troubling limitation of large language models: they often introduce details not found in the original documentation, resulting in fabricated specifications due to hallucination. This can ultimately yield factually inaccurate specifications.

\begin{figure}[!htbp]
\begin{tcolorbox}[
    title=\textbf{Example of Specification Fabrication},
    colback=pink!10,
    colframe=orange!60!brown!60,
    boxrule=2pt,
    width=\columnwidth]

\textbf{Text in the Document:}

``FOLL\_BEHAVE: controls whether follow points in the same direction as lead vehicle or towards it"

\textbf{Generated Specification:}

\textbf{Spec 1}: $\mathit{G}(\mathit{follow\_mode} \land \mathit{FOLL\_BEHAVE = 0} \\ \text{\quad\quad\quad\quad} \rightarrow \mathit{X}(\mathit{vehicle\_direction = leader\_direction}))$

\textbf{Spec 2}: $\mathit{G}(\mathit{follow\_mode} \land \mathit{FOLL\_BEHAVE = 1} \\ \text{\quad\quad\quad\quad} \rightarrow \mathit{X}(\mathit{vehicle\_direction = towards\_leader}))$
 
\end{tcolorbox}
\caption{Example of specification fabrication: The LLM fabricated incorrect parameter-behavior relationships not present in the source documentation.}
\label{fig:example_of_specification_hallucination}
\end{figure}

\Cref{fig:example_of_specification_hallucination} illustrates an instance of specification fabrication by the LLM, where it generated incorrect specifications for the FOLL\_BEHAVE parameter. Although the original documentation only provided a brief description of this parameter's function in controlling direction relative to a lead vehicle, the LLM incorrectly expanded on this limited information by fabricating temporal logic specifications that were not grounded in the source text. 

For instance, the LLM arbitrarily assigned binary values (0 and 1) to the FOLL\_BEHAVE parameter and associated these values with specific behaviors, even though the original documentation made no mention of such numerical values or their corresponding effects. Actually,  the FOLL\_BEHAVE parameter has more than two options, ``1" means the vehicle should face the lead vehicle, and ``2" means the direction of the vehicle is the same as the leader.
This highlights the potential for LLMs to generate incorrect specifications that may not reflect the actual system requirements.

\begin{longfbox}
\textbf{Answer to RQ1:}
Large language models demonstrate promising potential in the formal specification extraction task, achieving accuracy between 45.4\% and 51.7\% across three LLMs. However, the LLMs are limited by their tendency to oversimplify requirements and fabricate non-existent details, indicating substantial room for improvement in automated specification extraction tasks.
\end{longfbox}
 
\subsection{Insight}
Through the study result analysis, we identified two major limitations that hinder the effectiveness of LLMs in end-to-end specification extraction: (1) specification oversimplification and (2) fabrication. 
Firstly, when LLMs are used in single-query interactions, they often produce outputs that are limited in length and contain oversimplified specifications. This limitation becomes particularly apparent when dealing with documents that contain substantial amounts of non-trivial information, as it can lead to numerous contents being omitted.
Secondly, LLMs are also prone to incorporating factual errors into their generated outputs, which can ultimately yield incorrect specifications.

However, the accuracy and reliability of specification extraction are vital. The omission or inaccuracy of specifications can result in unidentified system vulnerabilities, thereby jeopardizing overall safety. Therefore, if we aim to leverage large models for automated specification extraction, we aspire to more comprehensively capture the existing specifications in the original text while reducing the influence of generated fabricated content.

We attribute these challenges to two intrinsic limitations of LLMs: (1) limited processing and expressive capabilities, and (2) hallucination. Firstly, they possess inherent limitations in their processing and expressive capabilities, making them ill-equipped to handle complex tasks like specification extraction. This task demands both significant processing power to handle large inputs and substantial expressive power to precisely convey multiple intricate specifications. Secondly, the possibility of LLM hallucination introduces a significant risk of generating fabricated content in their outputs, which reduces the reliability of generated specifications.

Motivated by human cognitive processes, we break down the complex task of specification extraction into two more manageable sub-tasks: (1) identifying sentences that convey specifications, which demands less expressive power, and (2) converting these specification sentences into Temporal logic formulas, which requires less processing power.

Based on that task decomposition design, we introduce an annotation-then-conversion methodology to extract specifications from documents. This method first involves annotating sentences in the document that contain specification information, where the LLM only needs to output the positional information of the relevant sentences, thereby reducing the demands on its expressive capabilities. Subsequently, each annotated sentence is converted into a precise temporal logic formula, which is a more manageable task that lowers the requirements of the LLM's processing capabilities.

Furthermore, our method yields $\text{(}sentence, specification\text{)}$ pairs, facilitating fact-checking and mitigating the impact of hallucinations. This decomposition and pairing strategy enables better utilization of large language models' strengths while minimizing the effects of their limitations.

\section{Methodology}
\label{sec:method}

\subsection{Overview}
\begin{figure}[!htbp]
\centering
\includegraphics[width=\linewidth]{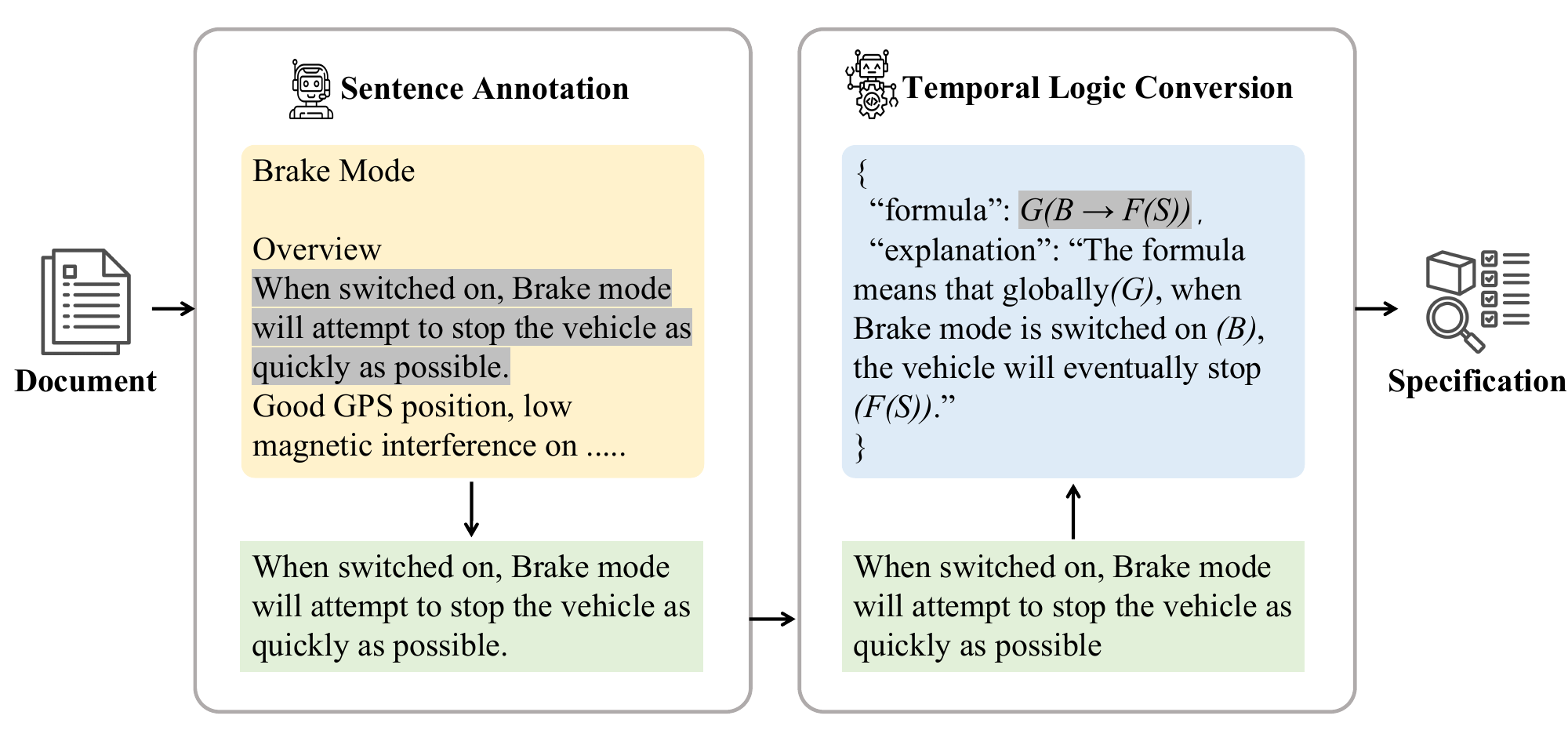}
\caption{Overview of the annotation-then-conversion method. Our method uses two LLMs: the Sentence Annotation agent takes in the document and identifies specification-related sentences, and the Temporal Logic Conversion agent takes in the annotated sentences and transforms these sentences into temporal logic formula. }
\label{fig:overview}
\end{figure}

Our method, as shown in the \Cref{fig:overview}, utilizes two specialized LLM agents to extract specifications from documents. The process involves two stages, where the first LLM agent is responsible for specification sentence annotation, identifying and tagging sentences in the document that contain specification information, and the second LLM agent performs temporal logic conversion, converting the annotated sentences into formal temporal logic formulas. By separating the tasks of annotation and conversion, our method enables more accurate and reliable specification extraction, allowing each agent to focus on its specific task, with the annotation agent concentrating on understanding natural language and identifying requirements, and the conversion agent applying its expertise in formal methods to generate well-formed temporal logic expressions.

\subsection{Sentence Annotation}

\begin{figure}[!htbp]
\centering
\begin{tikzpicture}[node distance=0.2cm]

\tikzstyle{instruction} = [rectangle, rounded corners, fill=instructionbg, text width=8cm, align=left, font=\small, inner sep=5pt]
\tikzstyle{objective} = [rectangle, rounded corners,  fill=codebg,  text width=8cm, align=left, font=\small, inner sep=5pt]
\tikzstyle{io} = [rectangle, rounded corners, fill=iobg, minimum width=2cm, align=left, font=\small, inner sep=5pt]
\tikzstyle{cot} = [rectangle, rounded corners, fill=cotbg,  text width=8cm, align=left, font=\small, inner sep=5pt]
\tikzstyle{req} = [rectangle, rounded corners, fill=reqbg,  text width=8cm, align=left, font=\small, inner sep=5pt]

\node[instruction] (roleplay) {
   \textbf{Roleplaying}\\
   You are an expert in the field of software engineering and are very skilled at text annotation.
};

\node[objective, below=of roleplay] (obj) {
  \textbf{Objective}\\
From the \verb+#@#@#@#@#@DOCUMENT#@#@#@#@#@+, annotate sentences
 as specifications if they convey the following information:
 
 \textcolor{blue}{//... We specify annotation rules here.}
};

\node[cot, below=of obj](cot) {
\textbf{Chain-of-Thought}\\
Let's go through this step by step to ensure we arrive at the correct answer.\\
STEPS:\\
\textcolor{blue}{//... We specify the steps of thought here.}
};

\node[req, below=of cot] (req) {
   \textbf{Requirment}\\
Specification sentences should be clear and specific; vague sentences should be excluded.
};

\node[io, below=of req.south west, text width=0.42\linewidth,anchor=north west] (input) {
\textbf{Input Format}\\
\begin{lstlisting}[basicstyle=\ttfamily\scriptsize]
{
  "sections": [
    {
    "id": ...,
    "sentences": [...]
    }, ...
  ]
}
\end{lstlisting}
};

\node[io, below=of req.south east, text width=0.42\linewidth, anchor=north east] (output) {
\textbf{Output Format}\\
\begin{lstlisting}[basicstyle=\ttfamily\scriptsize]
{
  "specifications": [
    {
    "section-id": "id", 
    "sentence-id": "id"
    }, ...
  ]
}
\end{lstlisting}
};
\end{tikzpicture}
\caption{Prompt for specification extraction. Our method incorporates a Chain-of-Thought methodology, which enables the LLM to decompose the annotation task into more tractable subtasks.}
\label{fig:prompt-spec-extract}
\end{figure}
Drawing inspiration from the prompt engineering practices used in the large language model research community \cite{shanahan2023role, wei2022chain, wu2023cheap}, we have designed a structured prompt that incorporates various elements to facilitate effective specification sentence annotation. As \Cref{fig:prompt-spec-extract} shows, these elements include roleplaying, objective, Chain-of-Thought, requirements for excluding undesired results, and input and output formats.

\subsubsection{Roleplaying}
Similar to the end-to-end method, our prompt template employs a roleplaying technique, enabling the LLM to assume the role of a domain expert in software engineering. This technique generally leads to better responses.

\subsubsection{Objective}
The primary objective of the sentence annotation agent is to accurately identify sentences that contain specification information pertinent to the software system. To accomplish this, we have defined four annotation rules that encapsulate common patterns of software specification information, outlined as follows:

\begin{itemize}
\item State Transition Requirements: The system must meet specific conditions before transitioning to a particular state. For example, the ArduPilot Brake mode document said \textit{``This mode requires GPS"} \cite{brake-mode}.

\item System Constraints: The system must adhere to specific ranges of important metrics.  For example, the ArduPilot Sport mode document said \textit{``The vehicle will not lean more than 45 degrees"} \cite{sport-mode}.

\item Expect Post Action: When the system enters a particular state, specific actions must be executed or important events must occur. For example, the ArduPilot RTL mode document said \textit{``When RTL mode is selected, the copter will return to the home location, or if rally points have been set up, the closet rally point"} \cite{rtl-mode}.

\item Expect State Change: The system must respond to user commands by executing specific actions or providing a response. For example, the ArduPilot Drift mode document said \textit{``If the pilot puts the throttle completely down the motors will go to their minimum rate and if the vehicle is flying it will lose attitude control and tumble"}   \cite{drift-mode}.
\end{itemize}

\subsubsection{Chain-of-Thought}
The Chain-of-Thought technique enables LLM to improve performance by further decomposing sentence annotation tasks into manageable subtasks that can be executed step by step. The process includes the following steps:
\begin{enumerate}
    \renewcommand{\labelenumi}{\arabic{enumi}.}
    \item Document Review: Thoroughly read the document to gain a comprehensive understanding of its content.
    \item Sentence Categorization: Analyze each sentence in context to determine whether it conveys information related to: ``State Transition Requirements", ``System Constraints", ``Expect Post Action" and ``Expect State Change".
    \item Specification Annotation: If a sentence falls into one of the four categories, annotate it as a specification sentence.
    \item JSON Formatting: Format all annotated sentences in JSON for further processing and analysis.
\end{enumerate}

\subsubsection{Requirements}
To exclude undesired output and improve the LLM's response, we specify requirements for the identified specification sentences in the prompt. We require the identified specification sentences to be clear and specific, making it easier to verify and validate them. Vague sentences should be excluded from the output, as they are unlikely to contain specification information and are difficult to verify.

\subsubsection{IO Format} 
A long output may exceed the context window of the LLM
and restrict the ability of the LLM.
Therefore,
to reduce the length of the output and let the LLM output as much as it can do, we assign every sentence within the document a unique pair of IDs 
$(section\_id, sentence\_id)$ and let the LLM only output the pair of IDs of the identified sentences. If the LLM outputs sentence IDs that are not in the document, we could simply reject them to reduce the influence of hallucination to a certain extent.
   
\subsection{Temporal Logic Conversion}
\begin{figure}[!htbp]
\centering
\begin{tikzpicture}[node distance=0.2cm]

\tikzstyle{instruction} = [rectangle, rounded corners, fill=instructionbg, text width=8cm, align=left, font=\small, inner sep=5pt]
\tikzstyle{objective} = [rectangle, rounded corners, fill=codebg, text width=8cm, align=left, font=\small, inner sep=5pt]
\tikzstyle{io} = [rectangle, rounded corners, fill=iobg, minimum width=2cm, minimum height=4.1cm, align=left, font=\small, inner sep=5pt]
\tikzstyle{cot} = [rectangle, rounded corners, fill=instructionbg, text width=8cm, align=left, font=\small, inner sep=5pt]

\node[instruction] (roleplay) {
   \textbf{Roleplaying}\\
    You are an expert in Temporal Logic (TL) with years of experience in formal verification and testing. 
};

\node[objective, below=of roleplay] (obj) {
   \textbf{Objective}\\
    Convert the given list of natural language sentences into Temporal Logic (TL) formulas. \\
    \textcolor{blue}{// We use the same operators as in the end-to-end task.}
};

\node[io, below=of obj, text width=8cm](io) {
\textbf{Example} \\
{\textbf{\textit{Input:}}}\\
Eventually, the system will reach a stable state and remain stable thereafter.

\bigskip

{\textit{\textbf{Output:}}}\\
\{\\
    \quad ``formula": $\mathit{F(RS} \land \mathit{G(S))}$, \\
    \quad ``explanation": ``Here, $\mathit{RS}$ represents the system reaching a stable state, and $\mathit{G(S)}$ ensures that stability($\mathit{S}$) is maintained indefinitely after that point."\\
  \}\\
};

\end{tikzpicture}
\caption{Prompt for TL conversion. We employ a few-shot learning technique to facilitate the LLM's conversion process by providing a concrete example of TL conversion.}
\label{fig:prompt-tl-conversion}
\end{figure}

We employ the LLM to translate annotated sentences into temporal logic formulas. As illustrated in \Cref{fig:prompt-tl-conversion}, we utilize a few-shot learning technique \cite{NEURIPS2020_1457c0d6} to instruct the LLM on how to perform this conversion. Few-shot learning enables LLMs to adapt to specific tasks, formats, or styles, thereby improving accuracy. However, since LLM's output can be creative, we need to ensure that the LLM's output strictly adheres to the temporal logic format. To achieve this, we provide a concrete example as part of the prompt template, which serves as a mapping guide for the LLM. Specifically, this example demonstrates how to translate natural language words into temporal logic operators, such as mapping ``eventually" to ``\text{F}" and ``remain" to ``\text{G}". By leveraging this example through few-shot learning, we improve the accuracy of the conversion process.

\section{Evaluation}
\label{sec:evaluation}
\subsection{Research Questions}
\noindent \textbf{RQ2}: Can the annotation-then-conversion technique improve the LLM's ability of formal specification extraction?

LLMs' limited processing and expressive power hinder their performance in end-to-end specification extraction tasks. To address this limitation, we propose a novel annotation-then-conversion method that aims to improve the accuracy of specification extraction. Our research question (RQ2) seeks to investigate whether this proposed method can outperform the traditional end-to-end method in terms of accuracy, thereby mitigating the negative impact of LLMs' limitations.

\noindent \textbf{RQ3}: How do LLMs perform on the task of temporal logic conversion compared to the state-of-the-art method in the pre-LLM era?

Natural languages are inherently ambiguous and imprecise, which poses a challenge for specification extraction. The annotation-then-conversion method relies on a crucial conversion step to transform informal and ambiguous sentences into clear and formal temporal logic specifications. However, if this conversion process is ineffective, it can lead to erroneous temporal logic specifications, ultimately compromising the overall performance of the specification extraction task. Given that converting natural language sentences into temporal logic specifications is a long-standing research problem, our research question (RQ3) aims to investigate whether state-of-the-art methods prior to the advent of LLMs can effectively facilitate the conversion of sentences conveying specifications into temporal logic formulas.

\subsection{Experimental Settings}
\subsubsection{Experiment 1: Evaluating the annotation-then-conversion method} Firstly, for each document file, We invoke LLM to annotate the sentences that convey specification information. To mitigate the influence of randomness, we invoke each LLM 3 times and evaluate the union of the result in 3 different invocations. Secondly, we invoke each LLM to convert the annotated sentences into temporal logic formulas. Finally, we ask for two experts to verify the specifications in temporal logic formulas. A specification is considered as right based on two criteria: First, it should exactly express what the corresponding document sentence has said. Second, it should match the specifications in the ground truth. Any temporal logic formula if its corresponding sentences convey information that doesn't fall in the ground truth dataset would be considered as wrong. By counting how many extracted specifications are valid, we could evaluate the effectiveness of our annotation-then-conversion method in terms of specification extraction.

\subsubsection{Experiment 2: Evaluating DeepSTL for Temporal Logic Conversion}We utilize DeepSTL \cite{he2022deepstl}, a state-of-the-art natural language to temporal logic conversion method, to generate temporal logic formulas from the extracted sentences. 

We conducted the experiment by running the DeepSTL temporal logic conversion tool on a machine with an openEuler release 20.03 LTS, featuring 80 CPU cores, 251GB RAM, 12GB GPU VRAM, and 870GB disk space. The input to the tool consisted of the sentences obtained from Experiment 1's sentence annotation agent. The resulting temporal logic formulas were then verified by two experts against the ground truth. This allowed us to compare the conversion results of DeepSTL with those of LLM, enabling an assessment of the impact of temporal logic conversion ability on overall performance.

\subsection{Results: Experiment 1}

\begin{table}[!ht]
    \centering
    \caption{Assessing the performance of the annotation-then-conversion method in extracting specifications from documents using LLMs. The structure is identical to \Cref{tab:end_to_end_result}.}
    \label{tab:annotation_then_conversion}
    \scriptsize
    \begin{tabular}{l|cc|cc|cc|c}
    \toprule
        \multirow{2}{*}{\textbf{Document}} & \multicolumn{2}{c|}{\textbf{Claude}} & \multicolumn{2}{c|}{\textbf{GPT-4o}} & \multicolumn{2}{c|}{\textbf{Llama}}& \textbf{Ground} \\
        \cline{2-7}
        & \textbf{r} & \textbf{w} & \textbf{r} & \textbf{w} & \textbf{r} & \textbf{w} & \textbf{Truth} \\ 
        \midrule 
        AP:Airmode & 7 & 1 & 4 & 0 & 6 & 0 & 8 \\ 
        AP:Auto & 22 & 1 & 18 & 1 & 19 & 9 & 29 \\ 
        AP:Brake & 4 & 2 & 5 & 1 & 6 & 2 & 8 \\ 
        AP:Circle & 21 & 3 & 18 & 1 & 19 & 4 & 25 \\ 
        AP:Drift & 12 & 3 & 7 & 1 & 9 & 2 & 14 \\ 
        AP:Flip & 7 & 3 & 8 & 2 & 4 & 4 & 9 \\ 
        AP:FlowHold & 5 & 3 & 2 & 1 & 5 & 2 & 8 \\ 
        AP:Follow & 11 & 2 & 8 & 1 & 3 & 3 & 12 \\ 
        AP:Guided & 16 & 6 & 10 & 5 & 19 & 8 & 27 \\ 
        AP:Heli\_Autorotate & 13 & 4 & 11 & 10 & 14 & 6 & 31 \\ 
        AP:Land & 9 & 1 & 6 & 3 & 7 & 1 & 11 \\ 
        AP:Loiter & 9 & 6 & 8 & 3 & 7 & 3 & 15 \\ 
        AP:PosHold & 8 & 1 & 3 & 2 & 4 & 1 & 11 \\ 
        AP:RTL & 28 & 1 & 15 & 5 & 39 & 5 & 44 \\ 
        AP:Simple & 10 & 2 & 5 & 3 & 14 & 3 & 19 \\ 
        AP:SmartRTL & 16 & 2 & 12 & 0 & 13 & 0 & 20 \\ 
        AP:Sport & 4 & 1 & 4 & 1 & 4 & 2 & 7 \\ 
        AP:Stabilize & 12 & 2 & 7 & 4 & 12 & 1 & 14 \\ 
        AP:SysID & 3 & 0 & 2 & 0 & 3 & 0 & 3 \\ 
        AP:Throw & 13 & 1 & 13 & 1 & 11 & 0 & 19 \\ 
        AP:Turtle & 8 & 1 & 9 & 0 & 7 & 1 & 12 \\ 
        PX4:Position & 18 & 4 & 8 & 2 & 11 & 5 & 20 \\ 
        PX4:Position Slow & 16 & 2 & 15 & 0 & 4 & 4 & 23 \\ 
        PX4:Altitude & 12 & 1 & 7 & 2 & 15 & 2 & 17 \\ 
        PX4:Stabilized & 11 & 4 & 8 & 0 & 12 & 4 & 16 \\ 
        PX4:Acro & 4 & 1 & 4 & 0 & 4 & 5 & 5 \\ 
        PX4:Hold & 11 & 1 & 6 & 1 & 5 & 2 & 13 \\ 
        PX4:Return & 18 & 1 & 15 & 3 & 15 & 2 & 22 \\ 
        PX4:Mission & 25 & 4 & 12 & 4 & 28 & 14 & 39 \\ 
        PX4:Takeoff & 8 & 1 & 7 & 0 & 7 & 1 & 11 \\ 
        PX4:Land & 12 & 0 & 7 & 0 & 10 & 0 & 13 \\ 
        PX4:Orbit & 15 & 1 & 10 & 3 & 10 & 4 & 27 \\ 
        AW:Blind Spot & 4 & 1 & 4 & 0 & 4 & 1 & 5 \\ 
        AW:Traffic Light & 9 & 1 & 9 & 0 & 9 & 1 & 9 \\ 
        AW:Detection Area & 8 & 0 & 7 & 0 & 7 & 2 & 8 \\ 
        AW:No Drivable Lane & 6 & 2 & 4 & 3 & 6 & 2 & 8 \\ 
        AW:Out of Lane & 17 & 6 & 12 & 9 & 10 & 4 & 21 \\  
        \midrule
        \textbf{Sum} & 432 & 73 & 310 & 72 & 382 & 110 & 603 \\ 
        \midrule
        \textbf{Accuracy} & \multicolumn{2}{c|}{71.6\%} & \multicolumn{2}{c|}{51.4\%} & \multicolumn{2}{c|}{63.3\%} & - \\ 
        \textbf{False Positive} & \multicolumn{2}{c|}{14.5\%} & \multicolumn{2}{c|}{18.8\%} & \multicolumn{2}{c|}{22.4\%} & - \\ 
        \bottomrule
    \end{tabular}
\end{table}

\Cref{tab:annotation_then_conversion} compares how well different LLMs (Claude, GPT-4o, and Llama) perform at extracting specifications using the annotation-then-conversion method. The result covers multiple software systems: ArduPilot (21 documents), PX4 (11 documents), and Autoware (5 documents). For each document, the table shows the number of correctly extracted specifications (\textbf{r}) and incorrectly extracted specifications (\textbf{w}), compared against a ground truth. The bottom rows summarize the overall performance with accuracy and false positive rates.

Our results demonstrate the efficacy of the annotation-then-conversion method for specification extraction, with the three LLMs successfully capturing a significant portion of the ground truth specifications. Specifically, the LLMs correctly extracted 432, 310, and 382 specifications out of a total of 603 in the ground truth data. This represents a 29.2\% improvement over the end-to-end method in terms of coverage, demonstrating the potential of the annotation-then-conversion method for automated specification generation. In terms of accuracy, Claude achieved a rate of 71.6\%, while GPT-4o and Llama achieved rates of 51.4\% and 63.3\%, respectively. These results highlight the effectiveness of the annotation-then-conversion method for specification extraction. 

Notably, the annotation-then-conversion method outperforms the end-to-end method when handling documents containing numerous specifications. A prime example is document \texttt{AP:RTL}, which contains 44 ground truth specifications, the highest among all documents. In this case, Claude extracted 28 correct specifications with only 1 error, while Llama extracted 39 correct specifications with 5 errors. These results demonstrate the annotation-then-conversion method's ability to effectively handle complex documents and accurately extract large numbers of specifications.

\begin{figure}[!htbp]
\begin{tikzpicture}
\begin{axis}[
    at={(-0.5cm,-0.5cm)}, 
    width=\linewidth,
    height=0.75\linewidth,
    ybar=5pt,
    enlarge x limits=0.2,
    ylabel={Percentage (\%)},
    ylabel style={yshift=-0.5cm}, 
    symbolic x coords={Claude,GPT-4,Llama},
    xtick=data,
    xtick style={draw=none}, 
    ytick style={draw=none}, 
    legend style={
        at={(0.5,1.05)},
        anchor=south,
        legend columns=4,
        font=\footnotesize,
        draw=none,
        /tikz/every even column/.append style={column sep=0.2cm},
    },
    ymin=0, 
    ymax=100, 
    bar width=10pt,
    ymajorgrids=true,
    grid style={dotted,gray!50},
    scaled y ticks=false,
    axis on top=false, 
    nodes near coords, 
    nodes near coords align={vertical}, 
    every node near coord/.append style={font=\tiny}
]

\addplot[ybar, fill=cyan, draw opacity=0] coordinates { 
    (Claude,51.7) (GPT-4,47.1) (Llama,45.4)
};
\addplot[ybar, fill=CarnationPink, draw opacity=0] coordinates { 
    (Claude,16.6) (GPT-4,22.6) (Llama,24.3)
};
\addplot[ybar, fill=LimeGreen, draw opacity=0] coordinates {
    (Claude,71.6) (GPT-4,51.4) (Llama,63.3)
};
\addplot[ybar, fill=Dandelion, draw opacity=0] coordinates {
    (Claude,14.5) (GPT-4,18.8) (Llama,22.4)
};
\legend{E2E-Acc, E2E-FP, ATC-Acc, ATC-FP}

\end{axis}
\end{tikzpicture}
\caption{Comparison accuracy and false positive between end-to-end(E2E) method and annotation-then-conversion(ATC) method. }
\label{fig:accuarcy_fp_compare}
\end{figure}
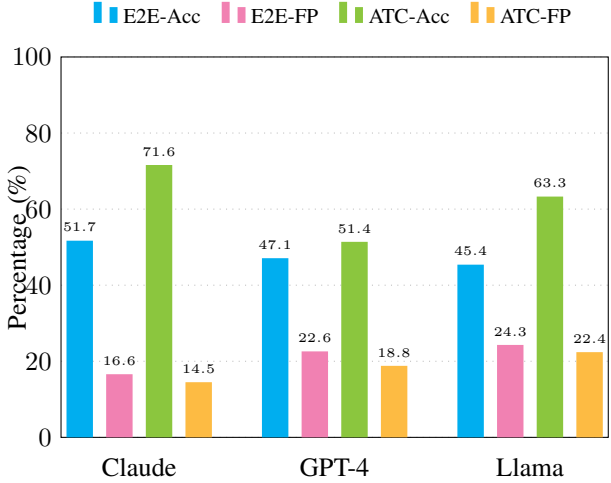

\Cref{fig:accuarcy_fp_compare} shows the difference of accuracy and false positive between end-to-end(E2E) method and annotation-then-conversion(ATC) method.  The results demonstrate that the annotation-then-conversion method significantly enhances specification extraction accuracy compared to end-to-end methods, with an average accuracy improvement of 14.0\%. For Claude, the accuracy increased from 51.7\% to 71.6\% when using the annotation-then-conversion method, representing a 19.9 percentage point improvement. Similarly, GPT-4o's accuracy increased from 47.1\% to 51.4\% (4.3 percentage point increase), while Llama showed substantial improvement from 45.4\% to 63.3\% (17.9 percentage point increase), indicating that the annotation-then-conversion method helps improve the precision of specification extraction.

Our method has also reduced in false positive rate. Claude's false positive rate decreased slightly from 16.6\% to 14.5\% (2.1 percentage point decrease), while GPT-4o's false positive rate decreased from 22.6\% to 18.8\% (3.8 percentage point decrease). Llama showed the highest false positive rates in both methods, decreasing from 24.3\% to 22.4\% with annotation(1.9 percentage point decrease)). While the method might still have some false positives for certain LLMs, the substantial gains in accuracy outweigh this drawback. More importantly, the method gives results in the format of  $\text{(}sentence, specification\text{)}$ pairs, which is an effective way for these false positives to be verified and fixed by human, thereby reducing their impact.

The results suggest that Claude outperforms other tested LLMs in terms of accuracy and false positive rates. With an accuracy rate of 71.6\% and a false positive rate of 14.5\%, Claude achieves a balance between precision and reliability.
This performance supports the effectiveness of the annotation-then-conversion method,  
where this task decomposition serves as an effective mechanism to unlock LLM's potential to extract specifications more accurately and reliably. 
The two-stage process proves to be a valuable methodology for extracting specifications, as confirmed by the experimental results.
\begin{longfbox}

\textbf{Answer to RQ2:}
The annotation-then-conversion method demonstrates superior coverage in formal specification extraction, resulting in a 29.2\% increase in the number of correct extracted specifications and an average accuracy improvement of 14.0\% compared to the end-to-end method. 
\end{longfbox}

\subsection{Results: Experiment 2}

\begin{table}[!ht]
    \caption{Specification extraction results after replacing the temporal logic conversion agent with the SOTA method DeepSTL in the pre-LLM era. The structure is identical to  \Cref{tab:end_to_end_result}.}
    \scriptsize
    \label{tab:deepstl_result}
    \begin{tabular}{l|cc|cc|cc|c}
    \toprule
        \multirow{2}{*}{\textbf{Document}} & \multicolumn{2}{c|}{\textbf{Claude}} & \multicolumn{2}{c|}{\textbf{GPT-4o}} & \multicolumn{2}{c|}{\textbf{Llama}} & \textbf{Ground} \\
        \cline{2-7}
        & \textbf{r} & \textbf{w} & \textbf{r} & \textbf{w} & \textbf{r} & \textbf{w} & \textbf{Truth} \\ 
        \midrule 
        AP:Airmode & 0 & 6 & 0 & 4 & 0 & 7 & 8 \\ 
        AP:Auto & 1 & 22 & 1 & 18 & 1 & 27 & 29 \\ 
        AP:Brake & 0 & 6 & 0 & 6 & 0 & 8 & 8 \\ 
        AP:Circle & 0 & 24 & 0 & 19 & 0 & 23 & 25 \\ 
        AP:Drift & 2 & 13 & 1 & 7 & 2 & 9 & 14 \\ 
        AP:Flip & 1 & 9 & 1 & 9 & 1 & 7 & 9 \\ 
        AP:FlowHold & 1 & 7 & 0 & 3 & 1 & 6 & 8 \\ 
        AP:Follow & 1 & 12 & 1 & 8 & 0 & 6 & 12 \\ 
        AP:Guided & 1 & 21 & 0 & 15 & 1 & 26 & 27 \\ 
        AP:Heli\_Autorotate & 5 & 12 & 5 & 16 & 3 & 17 & 31 \\ 
        AP:Land & 0 & 10 & 0 & 9 & 0 & 8 & 11 \\ 
        AP:Loiter & 3 & 7 & 4 & 7 & 3 & 12 & 15 \\ 
        AP:PosHold & 3 & 6 & 1 & 4 & 1 & 4 & 11 \\ 
        AP:RTL & 2 & 27 & 2 & 18 & 4 & 40 & 44 \\ 
        AP:Simple & 0 & 11 & 0 & 8 & 0 & 17 & 19 \\ 
        AP:SmartRTL & 0 & 18 & 0 & 12 & 0 & 13 & 20 \\ 
        AP:Sport & 1 & 4 & 0 & 5 & 1 & 5 & 7 \\ 
        AP:Stabilize & 0 & 14 & 0 & 11 & 0 & 13 & 14 \\ 
        AP:SysID & 0 & 3 & 0 & 2 & 0 & 3 & 3 \\ 
        AP:Throw & 4 & 9 & 4 & 10 & 3 & 8 & 19 \\ 
        AP:Turtle & 1 & 8 & 1 & 8 & 1 & 7 & 12 \\ 
        PX4:Position & 0 & 22 & 0 & 10 & 0 & 16 & 20 \\ 
        PX4:Position Slow & 0 & 18 & 0 & 15 & 0 & 8 & 23 \\ 
        PX4:Altitude & 3 & 10 & 0 & 9 & 3 & 14 & 17 \\ 
        PX4:Stabilized & 3 & 12 & 2 & 6 & 3 & 13 & 16 \\ 
        PX4:Acro & 1 & 4 & 1 & 3 & 1 & 8 & 5 \\ 
        PX4:Hold & 1 & 11 & 1 & 6 & 1 & 6 & 13 \\ 
        PX4:Return & 1 & 18 & 1 & 17 & 1 & 17 & 22 \\ 
        PX4:Mission & 2 & 27 & 1 & 15 & 3 & 39 & 39 \\ 
        PX4:Takeoff & 0 & 9 & 0 & 7 & 0 & 8 & 11 \\ 
        PX4:Land & 1 & 11 & 1 & 6 & 1 & 9 & 13 \\ 
        PX4:Orbit & 4 & 12 & 0 & 13 & 3 & 11 & 27 \\ 
        AW:Blind Spot & 0 & 5 & 0 & 4 & 0 & 5 & 5 \\ 
        AW:Traffic Light & 1 & 9 & 1 & 8 & 1 & 9 & 9 \\ 
        AW:Detection Area & 1 & 7 & 1 & 6 & 1 & 8 & 8 \\ 
        AW:No Drivable Lane & 0 & 8 & 0 & 7 & 0 & 8 & 8 \\ 
        AW:Out of Lane & 2 & 21 & 2 & 19 & 2 & 12 & 21 \\ 
        \midrule
        \textbf{Sum} & 46 & 453 & 32 & 350 & 42 & 457 & 603 \\
        \midrule
        \textbf{Accuracy} & \multicolumn{2}{c|}{7.6\%} & \multicolumn{2}{c|}{5.3\%} & \multicolumn{2}{c|}{7.0\%} & - \\ 
        \textbf{False Positive} & \multicolumn{2}{c|}{90.8\%} & \multicolumn{2}{c|}{91.6\%} & \multicolumn{2}{c|}{91.6\%} & - \\
        \bottomrule
    \end{tabular}
\end{table}

\Cref{tab:deepstl_result} presents the results of specification extraction when the temporal logic conversion agent in the annotation-then-conversion method is replaced with DeepSTL, a state-of-the-art method from the pre-LLM era. For clarity, we call the annotation-then-conversion method using two LLM agents ATC-LLM and call the replaced version ATC-DeepSTL. The table's structure is identical to that of the previous evaluation (Table \ref{tab:annotation_then_conversion}).
The results indicate that ATC-DeepSTL performs poorly in converting natural language sentences to temporal logic formulas. Notably, ATC-DeepSTL extracted significantly fewer specifications (46, 32, and 42) and its accuracy rates were remarkably low, ranging from 5.3\% to 7.6\%, whereas the ATC-LLMs achieved accuracy rates of up to 71.6\%. Moreover, ATC-DeepSTL's false positive rates were alarmingly high, exceeding 90\% in all cases. This suggests that ATC-DeepSTL not only struggles to correctly convert natural language sentences to temporal logic formulas but also produces a large number of incorrect formulas.

The subpar performance of ATC-DeepSTL can be attributed to DeepSTL's limitations in semantic understanding. Specifically, DeepSTL often struggles to accurately capture the logical relationships embedded in sentence semantics, even in short sentences. Furthermore, DeepSTL frequently fails to identify key variables and sometimes even generates random strings, leading to inaccurate temporal logic formulas.

For instance, when processing the sentence \textit{``This module is activated when there is traffic light in ego lane."}, DeepSTL generates the temporal logic formula ``{\fontfamily{lmtt}\selectfont always (Thismodule == activated)}" (``always" is equivalent to ``G"). The generated formula implies that the module is always activated, which is semantically inconsistent with the original sentence. This exemplifies the limitations of DeepSTL in comprehending the logical relationships inherent in sentence semantics.

Furthermore, DeepSTL exhibits significant limitations in identifying key variables. This is exemplified in the following generated temporal logic formula:

{\footnotesize\fontfamily{lmtt}\selectfont
"always(evhiclestoped==stoped\_stoped\_stop\_d\_stargin\_ds\\tae == etancerste ) until (not(erstacersta == stmoderste\_e\\ta) 
$\rightarrow$ (\_frot\_to\_t\_lineh== 
hlop\_d\_tstpo\_dsed\_ecinesa == ecedsahecedsledtase) until (esol"}

The generated formula contains numerous nonsensical variables and random strings, such as {\fontfamily{lmtt}\selectfont ecedsahecedsledtase}, which bear no resemblance to the original sentence's meaning. This highlights DeepSTL's inability to accurately identify and represent key variables, leading to the generation of meaningless temporal logic formulas.

\begin{longfbox}
\textbf{Answer to RQ3:}
LLM outperforms the traditional deep neural network-based tool on the task of converting natural language sentences to temporal logic formulas. Specifically, ATC-DeepSTL achieves an accuracy rate of up to 7.6\%, with false positive rates exceeding 90\%.
\end{longfbox}

\section{Threats to Validity}
The first concern is that the ground truth may be incomplete. To address this risk, we implemented an iterative process for establishing the ground truth. Specifically, we engaged experts in multiple rounds of specification identification to ensure that all relevant specifications were thoroughly extracted from the document.

The second concern is the potential for human error in manually verifying the extracted specifications in temporal logic. To mitigate this risk, we implemented a cross-validation process, where two experts review and verify each other's analysis results to ensure accuracy and consistency.

The third concern is related to output variability and time-Based Output Drift in LLMs \cite{sallou2024breaking}. We mitigate this by invoking each model three times and evaluating the union of outputs across trials.
\label{sec:limitation}

\section{Related Work}
\label{sec:related_works}

\noindent \textbf{Document Information Extraction.}
Document information extraction is a long-standing research area in NLP, focusing on extracting key information from various texts \cite{yang2022survey}, such as identifying rules in legal documents \cite{dragoni2016combining}. 
They have also applied in the software engineering field \cite{abad2018elica, sainani2020extracting, sudhi2023natural}, such as
extracting securities policies \cite{xiao2012automated},
requirement sentences \cite{haris2020automated},
and resource specifications \cite{zhong2009inferring}.
The methods used in this field can be broadly categorized into two types: rule-based and deep-learning-based methods.

After the appearance of LLM, LLM has been applied in extracting information from the document and gained significant improvement over the traditional method \cite{goel2023llms, sharma2023prosper, rejithkumar2024automated}.

\noindent \textbf{Generating temporal logic formulas from Natural language.}
The process of writing formal specifications in temporal logic has historically been a tedious and time-consuming endeavor. To mitigate this challenge, researchers have been actively investigating methods to generate temporal logic formulas from natural languages to simplify the process for users without extensive knowledge of temporal logic.
Their method can be classified into four categories: rule-based, deep-learning methods, and fine-tuning pre-trained model and model-based methods. The rule-based method employs parsers that utilize predefined rules to extract entities and their temporal relationships as intermediate representations, which are then translated into temporal logic formulas \cite{zhang2020automated,giannakopoulou2020formal,perez2022automated}.
In contrast, the deep-learning-based method relies on either training from scratch, where the translation is learned from a dataset of paired natural language and temporal logic formulas \cite{he2022deepstl,ge2023automtlspec,pan2023data}, or fine-tuning pre-trained model \cite{chen2023nl2tl}.
The LLM prompting-based method utilizes simple prompt engineering techniques that harness the power of large language models to perform the transformation \cite{manas2024tr2mtl,murphy2024guiding,nl2spec,aaai2023fc,liu2022lang2ltl,mavrogiannis2024cook2ltl}.

\section{conclusion}
\label{sec:conclusion}
Our study explored the feasibility of using Large Language Models for automated formal specification extraction from software documents. While LLMs showed promise, they struggled with oversimplification and fabrication of specifications. To address these limitations, we proposed a two-stage annotation-then-conversion method. This method resulted in a significant improvement in accuracy, averaging a 14.0\% increase, and a substantial rise in the number of extracted specifications, averaging a 29.2\% increase.
Our findings highlight the potential of LLMs for formal specification extraction, while also emphasizing the need for more effective methods. The proposed method offers a promising solution, and we believe it can significantly improve the accuracy and reliability of formal specification extraction in software engineering.
\section*{Acknowledgment}
We thank the anonymous reviewers for their insightful comments and suggestions.  This work was supported by the National Natural Science Foundation of China (Grant No. 62472100).

\clearpage
\bibliographystyle{IEEEtran}

\end{document}